\documentclass[manuscript]{aastex}
\usepackage{hyperref}
\usepackage{natbib}
\shorttitle{Ultra deep {\it AKARI} observations of Abell 2218}
\shortauthors{Hopwood et al.}

\begin{document}


\title{Ultra deep {\it AKARI} observations of Abell 2218: resolving the 15\,$\mu$m extragalactic background light }


\author{R. Hopwood\altaffilmark{1}, S. Serjeant\altaffilmark{1}, M. Negrello\altaffilmark{1}, C. Pearson\altaffilmark{1,2,3}, E. Egami\altaffilmark{4}, M. Im\altaffilmark{5}, J.-P. Kneib\altaffilmark{6}, J. Ko\altaffilmark{5}, H. M. Lee\altaffilmark{5}, M. G. Lee\altaffilmark{5}, H. Matsuhara\altaffilmark{7}, T. Nakagawa\altaffilmark{7}, I. Smail\altaffilmark{8}, T. Takagi\altaffilmark{7}}


\altaffiltext{1}{Department of Physics \& Astronomy, The Open University, Walton Hall, \\
Milton Keynes, MK7 6AA, UK}
\altaffiltext{2}{Space Science \& Technology Department, CCLRC Rutherford Appleton Laboratory, \\
Chilton, Didcot, Oxfordshire, OX11 0QX, UK}
\altaffiltext{3}{Department of Physics, University of Lethbridge, 4401 University Drive, Lethbridge, \\
Alberta T1J 1B1, Canada}
\altaffiltext{4}{Department of Astronomy, The University of Arizona, 933 North Cherry Avenue, \\
Rm. N204, Tucson, AZ 85721-0065, USA}
\altaffiltext{5}{Department of Physics \& Astronomy, FPRD, Seoul National University, \\
Seoul 151-742, Korea}
\altaffiltext{6}{OAMP, Laboratoire d'Astrophysique de Marseille, Traverse du Siphon, \\
13012 Marseille, France}
\altaffiltext{7}{Institute of Space and Astronautical Science, Japan Aerospace Exploration Agency,\\
Kanagawa 229-8510, Japan}
\altaffiltext{8}{Institute for Computational Cosmology, Durham University, South Road, \\
Durham, DH1 3LE, UK}

%Yoshinodai 3-1-1, Sagamihara, 


\begin{abstract}
We present extragalactic number counts and a lower limit estimate for the cosmic infrared background (CIRB) at 15\,$\mu$m from {\it AKARI} ultra deep mapping of the gravitational lensing cluster Abell 2218. These data are the deepest taken by any facility at this wavelength, and uniquely sample the normal galaxy population. We have de-blended our sources, to resolve photometric confusion, and de-lensed our photometry to probe beyond {\it AKARI}'s blank-field sensitivity. We estimate a de-blended 5\,$\sigma$ sensitivity of 28.7\,$\mu$Jy. The resulting 15\,$\mu$m galaxy number counts are a factor of three fainter than previous results, extending to a depth of $\sim$\,0.01\,mJy and providing a stronger lower limit constraint on the CIRB at 15\,$\mu$m of $1.9\,\pm$\,0.5\,nW\,m$^{-2}\,{\rm sr}^{-1}$.
\end{abstract}


\keywords{galaxies: clusters: individual (Abell 2218) --- galaxies: evolution --- infrared: galaxies}



\section{Introduction}
The cosmic infrared background (CIRB) is dominated by the dusty-emissions from star-forming galaxies, and therefore traces the dust-enshrouded star-formation over the history of the Universe. To interpret the CIRB in terms of galaxy formation and evolution models \citep[e.g.][]{2004ApJ...600..580G,2005MNRAS.358.1417P}, it is necessary to resolve the monochromatic backgrounds into their individual galaxies. There is a strong correlation between mid-IR and far-IR star-forming galaxies \citep{2001ApJ...556..562C,2002A&A...384..848E}, therefore galaxies responsible for the CIRB peak $\sim$140\,$\mu$m\,-\,200\,$\mu$m \citep{2006A&A...451..417D,2009Natur.458..737D} must also dominate the CIRB at shorter wavelengths, i.e., mid-IR $\lesssim$\,60\,$\mu$m. Recent results from {\it Spitzer}, {\it SCUBA} and {\it BLAST} data have shown, via stacking analysis of 24\,$\mu$m sources at longer wavelengths, that 24\,$\mu$m-selected populations account for the bulk of 70\,$\mu$m, 160\,$\mu$m and 250\,$\mu$m backgrounds and also dominate the 350\,$\mu$m, 450\,$\mu$m and 500\,$\mu$m backgrounds \citep{2006A&A...451..417D,2008MNRAS.386.1907S, 2009Natur.458..737D}. In contrast, at 850\,$\mu$m the 24\,$\mu$m population only resolves around one quarter of the background \citep{2008MNRAS.386.1907S}. Previous lower estimates of the CIRB at 15\,$\mu$m, from lensed ISOCAM data \citep[e.g.][hereafter MET03]{2003A&A...407..791M}, have successfully resolved the contribution from galaxies highly luminous in the IR, however the greater depth achieved by these data gives a more representative sample of galaxy populations.\\

A major challenge for deep IR observations is fluctuations from confusion noise, which presents a fundamental limit to blank-field surveys \citep{1974ApJ...188..279C}. Exploiting strong gravitational lensing offers a way to probe beyond the inherent blank-field confusion limit \citep{1997ApJ...490L...5S}. Where lensing increases the apparent surface area of a background field, the sources within that area are viewed at a lower number density in comparison to the un-lensed situation, and the preservation of surface brightness leads to amplified observed flux densities. Lensing therefore offers a twofold confusion-beating effect. Reduction in the observed area leads to source number-density depletion, and both area correction and flux correction are required to recover the true galaxy-number counts \citep[][]{1995astro.ph.11150B}. In addition to faint source confusion below the detection limit, a significant proportion of extractions from a confused image may be blends of two or more sources, so photometric de-blending is required \citep{2006MNRAS.371.1891R}. \\
\\
In this Letter we present new 15\,$\mu$m galaxy number counts and 15\,$\mu$m integrated light (IGL$_{15}$) estimate. In Section 2 we summarise the {\it AKARI} data and data reduction. A data analysis description is given in Section 3, and the results are presented in Section 4 and discussed in Section 5. \\
\\
Throughout this Letter we assume flat $\Lambda$CDM cosmology with $\Omega_{M}$\,=\,0.3 and $H_0$\,=\,70\,km\,s$^{-1}$\,Mpc$^{-1}$.

\section{Data acquisition and analysis}

\subsection{Data}

%% In a manner similar to \objectname authors can provide links to dataset
%% hosted at participating data centers via the \dataset{} command.  The
%% second curly bracket argument is printed in the text while the first
%% parentheses argument serves as the valid data set identifier.  Large
%% lists of data set are best provided in a table (see Table 3 for an example).
%% Valid data set identifiers should be obtained from the data center that
%% is currently hosting the data.
%%
%% Note that AASTeX interprets everything between the curly braces in the 
%% macro as regular text, so any special characters, e.g. "#" or "_," must be 
%% preceded by a backslash. Otherwise, you will get a LaTeX error when you 
%% compile your manuscript.  Special characters do not 
%% need to be escaped in the optional, square-bracket argument.

{\it AKARI} 15\,$\mu$m observations of A2218 were taken with the L15 filter of {\it AKARI}'s IRC ({\it AKARI:} \citet{2007PASJ...59S.369M}; IRC: \citet{2007PASJ...59S.401O}). The IRC has a wider field of view of 10\arcmin$\times$10\arcmin\ in comparison to the Infrared Array Camera (IRAC) aboard {\it Spitzer}, offering fuller coverage from 2\,$\mu$m\,-\,24\,$\mu$m. 19 pointings were acquired with the astronomical observation template IRC05, which is designed for deep observations and performs no dithering, however a nominal positional offset was applied in-between pointings.

%% In this section, we use  the \subsection command to set off
%% a subsection.  \footnote is used to insert a footnote to the text.


%% This section also includes several of the displayed math environments
%% mentioned in the Author Guide.

\subsection{Data reduction}

The data were reduced using the standard IRC-pipeline, version 20070912 \citep[][hereafter IRC-DUM]{ircman}. During the pipeline each pointing was divided into its constituent long and short exposures. The short exposures do not significantly add to the depth or quality of the final frame, so they are discarded. Each long exposure is 16.5 seconds, and up to 30 were average-combined per pointing, giving 19 reduced frames. We used optical data to register the frames astrometry.

%The pipeline failed to register positional information using 2MASS data, therefore we used optical data to register the frames.

The frames can be grouped based on the relative scanning direction of the IRC with A2218. An interval of roughly six months between the 10$^{th}$ and 11$^{th}$ pointings gave $\sim$180$\degr$ difference in the orientation of the first 10 frames (hereafter L15-A) and the final 9 frames (hereafter L15-B). This time interval led to an increase of bad pixels in the L15-B data, due to detector degradation. The L15-B data also suffer more severally from scattered light, a problem noted in the IRC-DUM. The scattered light is partially addressed by the pipeline, but remains an issue for several of the frames. A low-frequency sky noise is experienced by all post-pipeline frames. Combining the post-pipeline frames gives a significantly uneven background structure, which is detrimental to subsequent photometry. We therefore subtracted a median-filtered sky model, generated per frame, using a kernel width of 21.5\arcsec. A comparison of photometry taken for the image combined post-pipeline and the image combined after the additional sky-subtraction, showed good agreement at the bright end and a systematic shift at the faint end, attributable to the sky-structure present in the non-filtered image. We, therefore, concluded that the median-sky subtraction removes systematics associated with the extended sky-structure, without detriment to source photometry.

The final frames were average combined, giving an image (hereafter L15-image) with total integration time of 8460 seconds, full width at half maximum (FWHM) of the point-spread function (PSF) estimated at 5.96$\arcsec$ and a pixel scale of 2.39$\arcsec$. Figure 1 shows the post-pipeline combined image compared to the L15-image.

%\placefigure{fig_one}

\section{Analysis}

\subsection{Source extraction}

%% The equation environment will produce a numbered display equation.

A 5\,$\sigma$ source extraction was performed on the L15-image, with DAOFIND \citep{1987PASP...99..191S}. Combining the L15-A frames  and the L15-B frames into two `half-images' gave the means for a robust reliability check. Each source was examined individually in the `half-images', and those appearing at corresponding coordinates within both images were assumed to be real. 

\subsection{Sensitivity}
To estimate the map sensitivity we took aperture photometry at random positions, excluding the edges. An aperture radius of 5.96$\arcsec$ was used, and full flux densities were obtained using an aperture correction of 1.44 (See section 3.7) and the IRC-DUM ADU-to-$\mu$Jy conversion factor of 1.69\,$\mu$Jy ADU$^{-1}$. Fitting a Gaussian, with standard deviation 8.3\,$\mu$Jy, to the resulting distribution gave a 5\,$\sigma$ sensitivity estimate of 41.7\,$\mu$Jy. For the PSF-fitted catalogue (see Section 3.5) the sensitivity was estimated by comparing the input and output photometry for artificial sources introduced to the L15-image, giving a 5\,$\sigma$ sensitivity of 28.7\,$\mu$Jy.

%% The \notetoeditor{TEXT} command allows the author to communicate
%% information to the copy editor.  This information will appear as a
%% footnote on the printed copy for the manuscript style file.  Nothing will
%% appear on the printed copy if the preprint or
%% preprint2 style files are used.

%% The eqnarray environment produces multi-line display math. The end of
%% each line is marked with a \\. Lines will be numbered unless the \\
%% is preceded by a \nonumber command.
%% Alignment points are marked by ampersands (&). There should be two
%% ampersands (&) per line.

%% Putting eqnarrays or equations inside the mathletters environment groups
%% the enclosed equations by letter. For instance, the eqnarray below, instead
%% of being numbered, say, (4) and (5), would be numbered (4a) and (4b).
%% LaTeX the paper and look at the output to see the results.

%% This section contains more display math examples, including unnumbered
%% equations (displaymath environment). The last paragraph includes some
%% examples of in-line math featuring a couple of the AASTeX symbol macros.

\subsection{Multi-waveband counterparts}
We have multi-waveband coverage of A2218, taken by several facilities: {\it HST} WFPC2 F450, F606 and F814, Palomar 200 inch Hale {\it u', V, B, i'} and WHT's INGRID {\it Ks} and {\it J} \citep{2001MNRAS.323..839S,2001MNRAS.325.1571Z}; {\it Spitzer} IRAC Ch 1 to 4 and MIPS 24\,$\mu$m \citep[][in preparation]{Egami09}; {\it AKARI} S11 \citep{2009ApJ...695L.198K}. Figure 2 illustrates the A2218 coverage provided by this data set. 

To help identify blended sources within the extracted catalogue, a multi-waveband counterpart identification was performed. Potential counterparts for each L15-source were identified via a centroid search within a radius of 3.8$\arcsec$ (0.64\,$\times$\,FWHM) from the L15-centroid. For each source, a comparison of postage-stamp images across the available wavebands was performed to identify the main counterpart (brightest) plus subsidiary counterparts (less bright) and extra sources in the field not initially identified via the 5\,$\sigma$ extraction and within a radius of 18.0$\arcsec$.  

%\placefigure{fig_two}

\subsection{Field distortion}
A positional discrepancy between the L15-image and the counterpart images was identified during the counterparting process giving a spatially-varying PSF in the L15-image. Cubic polynomial coefficients were derived to map the L15-B frames onto the L15-A frame, and the resulting frames onto the Palomar and IRAC images. The L15-image was recombined and the empirical PSF was reconstructed, using PSTSELECT and PSF of the DAOPHOT package \citep{1987PASP...99..191S}, and showed no spatial variability. The source catalogue was then re-centred and the counterparting rerun. 

\subsection{De-blending and PSF-fitting}
Simultaneous PSF-fitting was performed on the L15-image using the full post-counterparting catalogue, which offers the benefit of positional priors. Constructing a reliably representative empirical PSF from a confused image is challenging, so we used all suitable sources available to statistically reduce noise in the PSF's tail. The PSF was refined following the iterative method outlined in the DAOPHOT2 manual, \citep{daophot2_2000}. The PSF radius was set to 17.9$\arcsec$, which collects approximately 100\% of the flux for non-extended sources according to the IRC-DUM and a plot of normalised pixel value as a function of radius, for sources in the L15-image. The resulting empirical PSF was used to CLEAN \citep{1974A&AS...15..417H} the L15-image with ALLSTAR \citep{1987PASP...99..191S}, giving a PSF-fitted source catalogue of 918 sources. The increase in the number of sources corresponds to a $\sim$\,40\% improvement in the completeness (see Section 3.6). 

\subsection{Completeness}
Two separate Monte Carlo completeness tests were run to represent the 5$\sigma$ catalogue and PSF-fitted catalogue. The first test used the established method of adding randomly-placed artificial sources to the L15-image, separated from the 5$\sigma$ catalogue, then performing an extraction on the results followed by aperture photometry of the extracted sources. The artificial sources were randomly scaled within defined flux density bins covering the source catalogue's flux range. For the first test the artificial source positions were generated with sufficient separation to avoid self-confusion. The 25$\arcsec$ minimum separation was derived by plotting normalised pixel values as a function of radius for bright well-separated sources. Twice the radius where the median pixel values disappear into the background was chosen. This test was repeated until around 20,000 sources per bin were achieved. The first test was adapted to represent the PSF-fitting of a photometrically-confused environment. Input positions were still generated randomly and kept at a distance from known source positions, although this limiting separation was reduced to 19$\arcsec$. A self-separation was imposed, but only to reject equal random positions. To reflect the use of positional priors and the re-centring carried out by ALLSTAR, the randomly-generated input positions were used as the ALLSTAR input rather than the extracted positions. This second test was repeated until around 30,000 sources per bin were achieved. \\
\\
Completeness was defined as the fraction of recovered sources per bin. The results of the first tests show that the L15-image is 10\%, 50\% and 90\% complete down to 20.2\,$\mu$Jy, 30.7\,$\mu$Jy and 46.8\,$\mu$Jy, respectively. For the second test, the L15-image is 10\%, 50\% and 90\% complete down to 12.2\,$\mu$Jy, 20.0\,$\mu$Jy and 31.5\,$\mu$Jy, respectively.

\subsection{Multi-waveband photometry}
{\it HST} photometry was obtained from the published catalogue of \citet{2001yCat..73230839S}. IRAC aperture photometry was taken with an aperture radius of 2.44$\arcsec$ and an annulus of radii 14.6$\arcsec$ and 24.4$\arcsec$, and the published IRAC aperture corrections were applied. For the remaining counterpart images, aperture photometry was taken and a growth-curve aperture correction method was employed \citep{1989PASP..101..616H,1990PASP..102..932S}. For each image a median growth-curve was empirically constructed using aperture photometry taken for bright and well-separated sources, with concentric apertures of increasing radii. The {\it u'} to {\it Ks} images were better represented by two growth-curves, one for point-like sources and the other for elliptical sources, which are not significantly extended. Aperture corrections were chosen on a source-by-source basis to minimise contamination from neighbours. For the L15-image an aperture correction for a radius of 5.96$\arcsec$ was derived, using the empirical PSF and a comparison of the PSF-fitted photometry and aperture photometry.

\subsection{Photometric redshifts}
Two codes that utilise a minimum $\chi^2$ spectral energy distribution (SED) fitting method were applied to estimate photometric redshifts for L15-sources with photometry coverage in four or more filters, shortwards of 11\,$\mu$m. EaZy \citep{2008ApJ...686.1503B} is suitable for data sets with few or biased spectroscopic redshifts (z$_{spec}$), such as the z$_{spec}$ available for the L15-catalogue, which are mainly biased at the cluster redshift of 0.18. The EaZy theoretical SED templates are based on semi-analytical models, and a linear combination of templates can be fitted simultaneously. IRAC photometry was included due to EaZy's ability to fit photometry up to IRAC CH4, however this is dependent on redshift. EaZy gives the option to apply priors aimed at breaking the template colour degeneracies seen with increasing redshift. Our spectra were also fitted using the photometric code of \citet[][hereafter N09]{2009MNRAS.394..375N}. This code is uniquely optimised for fitting mid-to-far-infrared polycylic aromatic hydrocarbon (PAH) and silicate features seen in starburst SEDs. For sources with strong mid-infrared PAH features, the comparison of N09 and EaZy redshifts was consistent with a slope of 1. A robust catalogue of photometric redshifts was constructed using a visual triple-check per source to reject unreliable estimates. The best SED fits from EaZy and N09 and the source morphology were visually compared, in context of the redshift estimate and probability of the minimum $\chi^{2}$. The redshift catalogue was constructed primarily from EaZy estimates. For sources with pronounced mid-infrared features and reliable EaZy and N09 estimates, not in agreement within their 1$\sigma$ errors, the N09 estimates were used when clearly providing additional constraint from fitting to mid-IR photometry. Cluster members were identified from spectroscopic redshifts or during the triple comparison, from their typical SED and elliptical morphology. Cluster members represent 16\% of the total catalogue, including all significantly extended sources. All cluster members were subsequently removed. 31\% of the remaining catalogue are without a redshift estimate, either due to a lack of multi-wavelength coverage or unreliable photometric estimate. For these sources a redshift of 1.04\,$\pm\,0.67$ was assigned, which is the median of the redshift catalogue with 1$\sigma$ errors. Substituting a value of 2.0 or 3.0, in place of the median redshift value, gave no significant difference for the resulting number counts.  

\section{Results}

\subsection{A2218 mass model}

Magnification corrections ($\mu$) were obtained using LENSTOOL \citep{2007NJPh....9..447J}, which required as input a mass model of A2218 and the positions and redshifts for all sources beyond the cluster distance. A pseudo isothermal elliptical mass model is assumed for the mass distribution of the A2218 cluster members. Strongly lensed arcs and arclets are used to constrain the mass distribution and total mass of the cluster \citep{2007arXiv0710.5636E, 1996ApJ...471..643K}. Spectroscopy of arclets has been used to test this model for reliability \citep{1998MNRAS.295...75E}. 

\subsection{Galaxy number counts}

Flux densities ($S$) were corrected prior to counting as $S_{{\rm true}}=\frac{S_{{\rm obs}}}{\mu} $. Corrections for depletion and incompleteness were applied to individual sources during counting, assuming the relation $n_{{\rm true}}\,=\,\frac{\mu}{C(S_{{\rm obs}})}$, where $n_{true}$ is the true number of sources and the completeness ({\it C}) is a function of $S_{{\rm obs}}$ (rather than $S_{{\rm true}}$). De-lensed number counts over bin {\it dS} are then obtained as $\frac{dN}{dS}\,=\,\Sigma n_{{\rm true}}$. The amplification ($\frac{1}{\mu}$) distribution ranges from 1.0 to 24 and has a median of 1.2, which reflects the wide area of A2218 covered and the decrease of amplification as a function of radius from the centre of the core. The $\mu$ distribution obtained was not found to change significantly with the variation of redshifts within the L15-redshift distribution.\\
\\
Figure 3 shows our Euclidean-normalised differential number counts, in comparison to a compilation of previous work and predictions based on galaxy evolution models. The median completeness per bin is 100\% down to the faintest three bins, which have median completeness corrections of 26\%, 70\% and 96\% respectively. Our L15-counts extend the faint end of observed counts down to $\sim$\,0.01\,mJy, which is a factor of 3 fainter in comparison to the ISOCAM \citep{1996A&A...315L..32C} lensing survey counts of MET03. Below 0.2\,mJy the L15-counts present a steep sub-Euclidean slope of -1.6, which agrees with the faint slope of \citet{1999A&A...351L..37E}. The no-evolution model is strongly excluded by all available data and there is a general consensus on a significant evolutionary bump, which peaks around 0.2-0.4\,mJy. The comparably steep slope of the L15-counts brighter than the 'bump' is the result of de-blending. The Pearson 2007/2010 model (hereafter P10) predicts that the populations dominating the 15\,$\mu$m counts `bump' are starbursts (L$\,<\,$10$^{11}$\,L$_\odot$) and luminous infrared galaxies (LIRG; 10$^{11}$\,L$_\odot\,<\,$L$\,<\,$10$^{12}$\,L$_\odot$), with redshift distributions peaking at z\,=\,0.5 and z\,=\,1.2, respectively. At a mean redshift of 0.8, in the bump, S$_{15}\,=\,0.3$\,mJy corresponds to $\sim$\,4$\times$10$^{11}\,$L$_\odot$ for an M82 SED.

%\placefigure{fig_three}

\subsection{Bootstrapping}

Confidence intervals for the differential counts were derived by bootstrapping within the photometric and redshift errors, as these are the dominant source of uncertainties for the counts. The L15-flux population was re-sampled without bias for the lensed-source catalogue. Each sample was randomly assigned flux densities and redshifts within the respective 3\,$\sigma$ errors, and de-lensed with recalculated magnification corrections. 30,000 re-sampled populations were generated and differential counts were taken for each. Confidence intervals were calculated using the median and standard deviation of the resampled counts. The grey shaded region of Figure 3 shows the resulting 95\% confidence interval. The divergence of the bootstrapping below 0.02\,mJy shows that the upturn of the counts in the final bin is not statistically significant. The differential number counts and the bootstrapped standard deviation are presented in Table 1.
 
 \subsection{15\,$\mu$m integrated galaxy light}
The IGL$_{15}$, a lower limit for the CIRB$_{15}$, can be obtained by integrating the flux per unit area for the corresponding monochromatic number counts. The differential contribution to the IGL$_{15}$ is given by \\
\hspace*{3cm}\begin{math}\displaystyle\frac{d{\rm IGL}}{dS}\,=\,\frac{dN}{dS}(\frac{S}{10^{20}})\nu_{15}\end{math} \hspace{5cm} \citep[e.g.][]{2002A&A...384..848E}\\
\\
where 1\,mJy\,=\,$\frac{1}{10^{20}}$\,nW\,m$^{-2}$\,sr$^{-1}$ and $\nu_{15}$ is the frequency of the 15\,$\mu$m photons.

Using the counts of \citet[][re-calibrated following \citet{2002MNRAS.337.1043V}]{2000MNRAS.316..768S}, \citet{2002MNRAS.335..831G,1993ApJS...89....1R} and the L15-counts, giving a flux range of 0.01\,mJy-10,000\,mJy, we estimate IGL$_{15}$\,=\,1.9\,$\pm$\,0.5\,nW\,m$^{-2}$\,sr$^{-1}$. The lensed fields observed by ISOCAM (including A2218) produced 15\,$\mu$m counts down to 0.03\,mJy and a lower-limit estimation for the IGL$_{15}$ of 2.7\,$\pm$\,0.62\,nW\,m$^{-2}$\,sr$^{-1}$ (MET03, and references therein). These estimates agree, within the errors. Using the lower flux limit of MET03, and our methodology, gives an IGL$_{15}$ of 1.6\,$\pm$\,0.38\,nW\,m$^{-2}\,$sr$^{-1}$, which is marginally consistent ($\sim$\,2\,$\sigma$) with the MET03 estimate. The P10 model provides an excellent fit to the whole of our data. Integrating over the full flux range of the P10 model gives a predicted IGL$_{15}$ of 2.3\,nW\,m$^{-2}$\,sr$^{-1}$. If we assume the shape (but not the normalisation) of the P10 counts, we derive a slightly better estimate of IGL$_{15}$\,=\,2.0\,$\pm$\,0.4\,nW\,m$^{-2}$\,sr$^{-1}$ at $\geq$\,0.01$\,$mJy (see Figure 4). In comparison to the MET03 result, which resolved $\sim\,70\%$ of the CIRB$_{15}$ into individual galaxies, we are $\times$3 deeper and resolve $87\%\,\pm\,13\%$ and, whereas the {\it Infrared Space Observatory (ISO)} surveys mainly sample galaxies with luminosities $\geq$\,LIRG, we are probing the more normal galaxy populations. 

\section{Discussion and conclusions}
From our de-blended and de-lensed 15\,$\mu$m counts we have derived an IGL$_{15}$ estimate of 2.0\,$\pm$\,0.4\,nW\,m$^{-2}$\,sr$^{-1}$, down to $\sim$\,0.01\,mJy. We conclude that, with respect to the P10 model, the {\it AKARI} 15\,$\mu$m data are consistent with having resolved the whole of the predicted IGL$_{15}$. Assuming no radical change between the IR SED of high-redshift galaxies and those resolved at 15\,$\mu$m with median redshift of 1.0 \citep{2002A&A...384..848E}, then the galaxies resolved by these data represent the bulk of galaxies dominating CIRB peak. 

Figure 4 suggests that in order to resolve 100\% of the CIRB$_{15}$, future observations need to probe depths in the region of 1 magnitude fainter than the sensitivity limit achieved by this survey, down to at least {\it S}$_{15}$\,=\,1\,$\mu$Jy. The first possible direct measurement constraints of the CIRB$_{15}$ will come from {\it JWST} or {\it SPICA} \citep{2006SSRv..123..485G,2004AdSpR..34..645N}. 

15\,$\mu$m stacking analysis of {\it Herschel}/SPIRE and PACS A2218 data, will address how representative 15\,$\mu$m selected galaxies are of the galaxy populations responsible for the CIRB at its peak.


%% Included in this acknowledgments section are examples of the
%% AASTeX hypertext markup commands. Use \url without the optional [HREF]
%% argument when you want to print the url directly in the text. Otherwise,
%% use either \url or \anchor, with the HREF as the first argument and the
%% text to be printed in the second.

\acknowledgments

We thank the anonymous referee for insightful and constructive comments. We thank the Great Britain Sasakawa Foundation for support with grant number 3108, the Royal Society for an international travel grant and the Science and Technology Facilities council, grant D/002400/1 and studentship SF/F005288/1. This research is based on observations with {\it AKARI}, a JAXA project with the participation of ESA. MI and JK were supported by the Korea Science and Engineering Foundation(KOSEF) grant No. 2009-0063616, funded by the Korea Government(MEST). IRS acknowledges support from STFC.


%% in parentheses, after the keyword, but they will not be verified.

{\it Facilities:}\facility{ AKARI (JAXA/ISAS)}.

\begin{figure*}[!htbp]
\begin{center}
\includegraphics[width=0.7\textwidth]{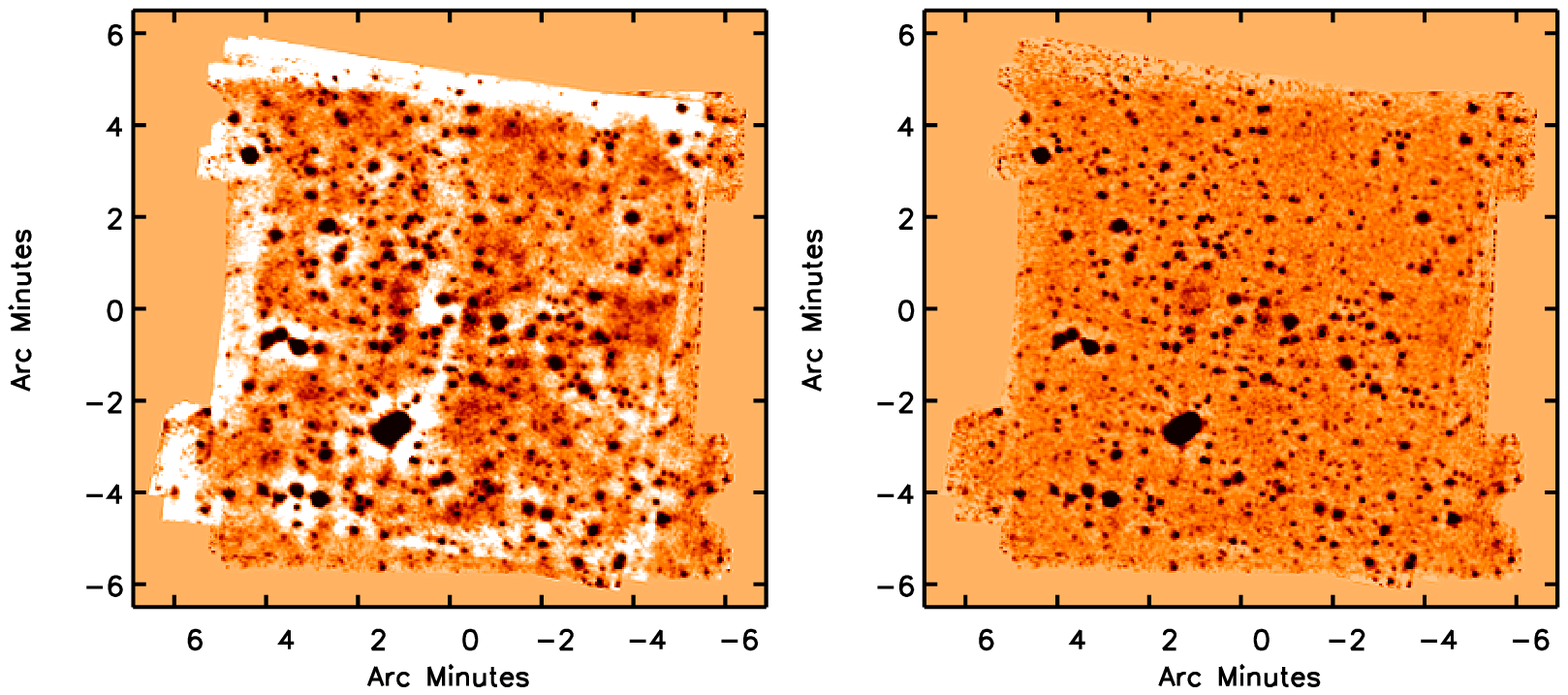}
\end{center}
\caption{Comparison between the L15-image combined post-pipeline (left) and the L15-image combined after the further reduction (right). The same pixel scaling was used to plot both images. The low-frequency background structure evident in the post-pipeline image is successfully removed by the additional reduction.}
\label{fig_one}
\end{figure*}

\begin{figure*}[!htbp]
\begin{center}
\includegraphics[width=0.7\textwidth]{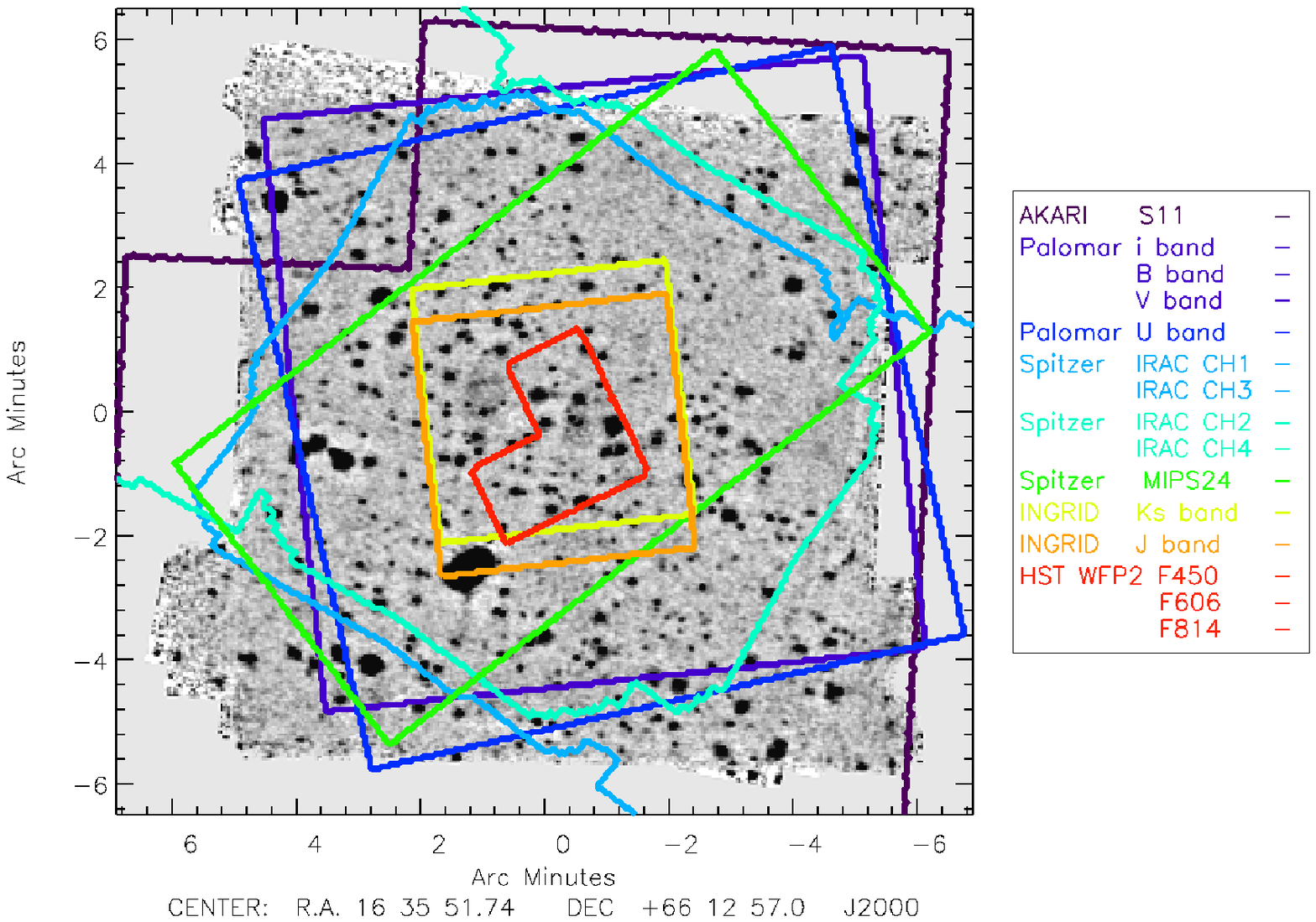}
\end{center}
\caption{L15-image overlaid with contours indicating the multi-wavelength coverage of A2218.}
\label{fig_two}
\end{figure*}

\begin{figure*}[!htbp]
\begin{center}
\includegraphics[width=0.85\textwidth]{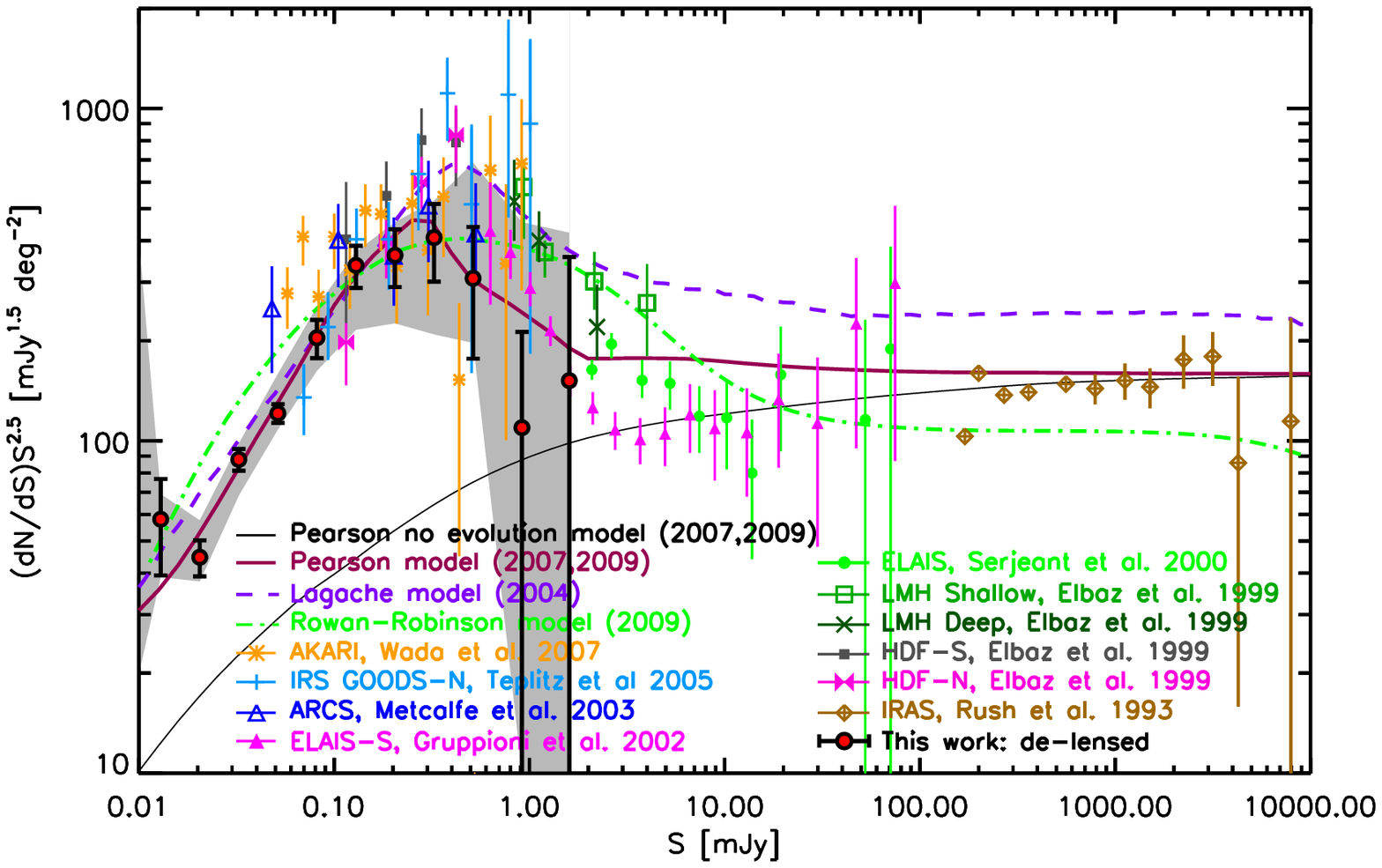}
\end{center}
\caption{Euclidean-normalised differential galaxy number counts. The L15-de-lensed counts are compared to the P10 model counts and no-evolution model, \citep{2007AdSpR..40..605P,pearson09}, the \citet{2009MNRAS.394..117R} model, the \citet{2004ApJS..154..112L} model, {\it IRAS} counts from \citet{1993ApJS...89....1R} (shifted from 12\,$\mu$m), {\it ISO} counts from \citet{1999A&A...351L..37E}, \citet[][re-calibrated following \citet{2002MNRAS.337.1043V}]{2000MNRAS.316..768S}, \citet{2002MNRAS.335..831G} and \citet{2003A&A...407..791M}, IRS counts from \citet{2005ApJ...634..128T} and {\it AKARI} counts from \citet{2007PASJ...59S.515W}. The grey shaded area represents the 2$\sigma$ bootstrapped confidence interval for the L15-counts. Note the good agreement of the faint end of the L15-counts with the Pearson and Lagache models.}
\label{fig_three}
\end{figure*}

\begin{figure*}[!htbp]
\begin{center}
\includegraphics[width=0.85\textwidth]{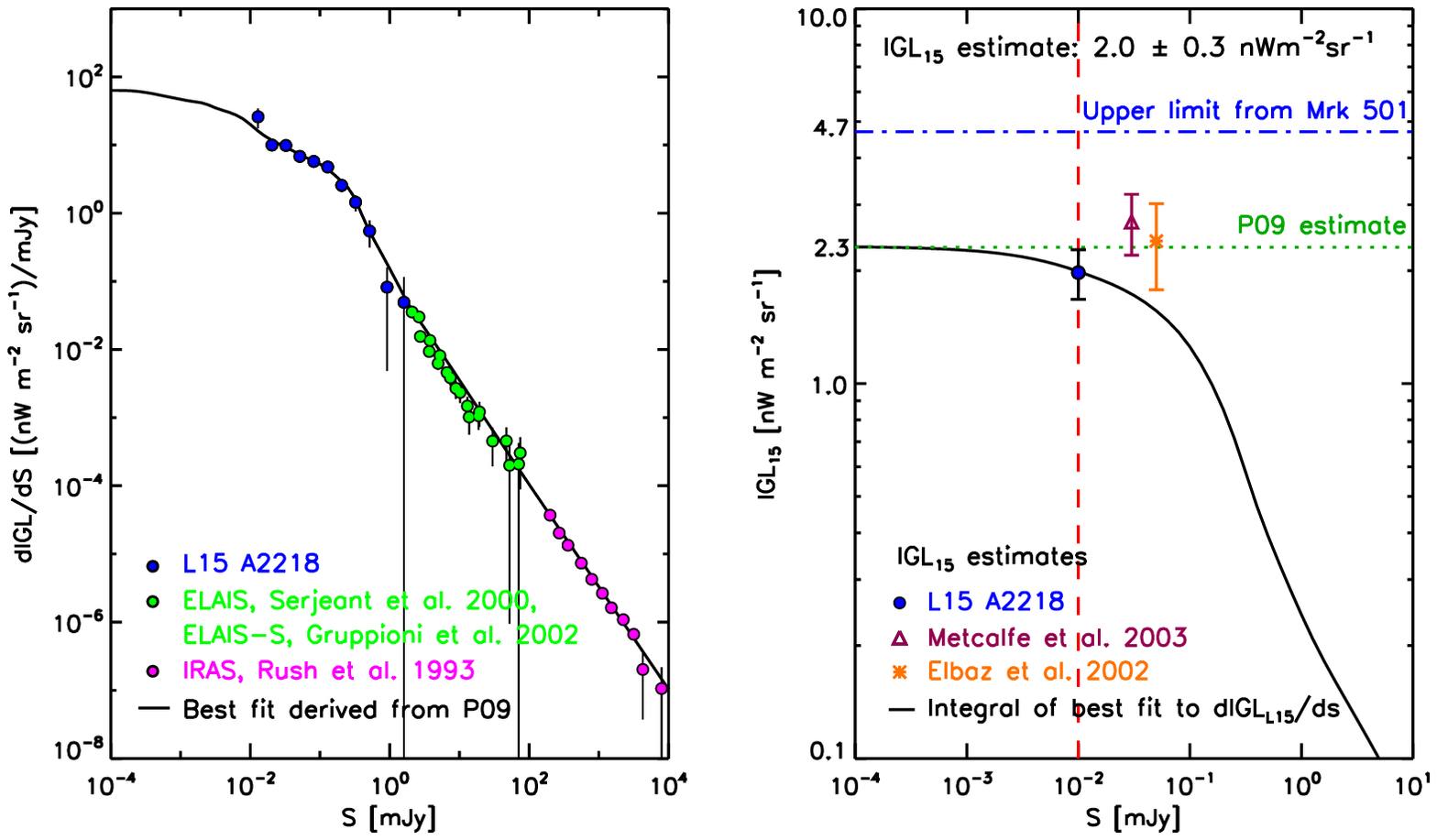}
\end{center}
\caption{Differential contribution to the IGL$_{15}$ as a function of flux density (left). Data shown are the L15-data, data from \citet[][]{2000MNRAS.316..768S,2002MNRAS.335..831G} and \citet[][]{1993ApJS...89....1R}. The black line represents the best fit of the P10 model to the L15-data. IGL$_{15}$ estimates as a function of flux density (right). The L15-estimate is limited to 0.01\,mJy, illustrated by the dashed red line. The \citet{2002A&A...384..848E} and \citet{2003A&A...407..791M} estimated limits are 0.05 and 0.03\,mJy respectively. The IGL$_{15}$ upper limit was derived from $\gamma$ ray emission of Mrk 501 \citep{2001A&A...371..771R}.
}
\label{fig_four}
\end{figure*}

\begin{deluxetable}{ccccc}
\tabletypesize{\footnotesize}
\tablecaption{Lensed ($\frac{dN(S_{\mathrm{ob}})}{dS_{\mathrm{ob}}}$) and De-lensed ($\frac{dN(S_{\mathrm{true}})}{dS_{\mathrm{true}}}$) Differential Number Counts, corrected for Incompleteness, and the Associated Bootstrapped Median Number Counts and Standard Deviation for the De-lensed Counts.}
\tablewidth{0pt}
\tablehead{
\colhead{Bin$_{L}$\,(mJy)} & \colhead{$\frac{dN(S_{\mathrm{ob}})}{dS_{\mathrm{ob}}}$({\tiny{mJy$^{-1}$deg$^{-2}$}})}& \colhead{$\frac{dN(S_{\mathrm{true}})}{dS_{\mathrm{true}}}$({\tiny{mJy$^{-1}$deg$^{-2}$}})}&\colhead{$\frac{dN(S_{\mathrm{true}})}{dS_{\mathrm{true}}}_{\mathrm{bootstrap}}$({\tiny{mJy$^{-1}$deg$^{-2}$}})}&\colhead{$\sigma_{\mathrm{bootstrap}}$({\tiny{mJy$^{-1}$deg$^{-2}$}})}
}
\startdata

1.00E-02 & (4.37$\pm$0.98)E+06 & (3.06$\pm$0.99)E+06 & 2.76E+06 & 3.54E+05 \\
1.59E-02 & (8.60$\pm$0.58)E+05 & (7.44$\pm$0.92)E+05 & 7.71E+05 & 7.13E+04\\
2.51E-02 & (3.61$\pm$0.81)E+05 & (4.63$\pm$0.35)E+05 & 4.27E+05 & 3.28E+04\\
3.98E-02 & (2.23$\pm$0.12)E+05 & (2.02$\pm$0.13)E+05 & 2.16E+05 & 1.80E+04\\
6.31E-02 & (9.42$\pm$9.47)E+04 & (1.08$\pm$0.14)E+05 & 1.06E+05 & 1.03E+04\\
1.00E-01 & (4.44$\pm$0.52)E+04 & (5.62$\pm$0.81)E+04 & 4.73E+04  & 5.70E+03\\
1.59E-01 & (1.93$\pm$0.27)E+04 & (1.90$\pm$0.38)E+04 & 1.73E+04 & 2.73E+03\\
2.51E-01 & (7.17$\pm$1.31)E+03 & (6.81$\pm$1.79)E+03 & 5.92E+03 & 1.22E+03\\
3.98E-01 & (2.56$\pm$0.62)E+03 & (1.62$\pm$0.69)E+03 & 2.24E+03 & 6.00E+02\\
6.31E-01 & (2.47$\pm$1.23)E+02 & (1.37$\pm$1.29)E+02 & 2.65E+02 & 1.37E+02\\
1.20E+00 & (0.00$\pm$0.00)E+00 & (4.68$\pm$6.35)E+01 & 4.39E+01 & 4.01E+01\\
\enddata
\tablecomments{ \,Lower bin limits are given}
\end{deluxetable}
\label{counts_table}

\end{document}